\documentclass[11pt]{article}
\usepackage{hyperref}
\pdfoutput=1
\begin{document}
\title{Counter Rotating Open Rotor Animation using Particle Image Velocimetry}
\author{
  E.W.M. Roosenboom, A. Schroeder, R. Geisler, D. Pallek and J. Agocs
  \vspace{6 pt}\\
   \small DLR - German Aerospace Center,\\
   \small Institute of Aerodynamics and Flow technology,\\
   \small Bunsenstr. 37073 Goettingen, Germany\\
  \and\\
  K.-P. Neitzke
  \vspace{6 pt}\\
  \small Airbus Operations GmbH,\\
  \small Wind Tunnel Test Engineering,\\
  \small Airbus Allee 1, 28199 Bremen Germany
 }
\date{}
\maketitle
\begin{abstract}
This article describes the two accompanying fluid dynamics videos for the ``Counter rotating open rotor flow field investigation using stereoscopic Particle Image Velocimetry'' presented at the 64th Annual Meeting of the APS Division of Fluid Dynamics in Baltimore, Maryland, November 20-22, 2011.
\end{abstract}
\section{Introduction}
Both videos (\href{file://anc/anim_dfd_HQ.avi}{High quality video} and \href{file://anc/anim_dfd_LQ.avi}{Low quality video}) show the vorticity evolution at a counter rotating open rotor configuration acquired with stereoscopic Particle Image Velocimetry in phase locked sense. The color coded vectors represent the convection velocity and only every fourth vector and second vector are shown in horizontal and vertical direction, respectively. For each measurement 500 instantaneous PIV acquisitions have been obtained. Each frame is subsequently ordered in bins of 3 deg.
\renewcommand{\thefootnote}{}\footnotetext{Copyright \copyright  2011 Airbus S.A.S., DLR - German Aerospace Center and the authors}
\end{document}